# Ion induced segregation in gold nanostructured thin films on silicon

J. Ghatak and P. V. Satyam[*]

Institute of Physics, Sachivalaya Marg, Bhubaneswar 751005, India

**Abstract:**

We report a direct observation of segregation of gold atoms to the near surface regime due to 1.5 MeV $Au^{2+}$ ion impact on isolated gold nanostructures deposited on silicon. Irradiation at fluences of $6\times10^{13}$, $1\times10^{14}$ and $5\times10^{14}$ ions $cm^{-2}$ at a high beam flux of $6.3\times10^{12}$ ions $cm^{-2}$ $s^{-1}$ show a maximum transported distance of gold atoms into the silicon substrate to be 60, 45 and 23 nm, respectively. At a lower fluence ($6\times10^{13}$ ions $cm^{-2}$) transport has been found to be associated with the formation of gold silicide ($Au_5Si_2$). At a high fluence value of $5\times10^{14}$ ions $cm^{-2}$, disassociation of gold silicide and out-diffusion lead to segregation of gold to defect - rich surface and interface region.

Metal-semiconductor interfaces and the diffusion of metal into Si are very important in the semiconductor technology and in particular, Au and Si are the two commonly used elements in semiconductor device production. The diffusion of noble metals in crystalline Si ($c$-Si) has been explained using "Frank-Turnbull" mechanism [1] and "kick-out" mechanism [2]. In case of amorphous silicon, the Au diffusion was explained in terms of direct diffusion in which the diffusing Au atoms are temporarily trapped by different kind of vacancy-liked effect [3, 4]. Past investigations of Au-implanted silicon have been suggested about the gold segregation as a result of its expelling from the recrystallized amorphous layer during thermal and ion beam annealing [5]. It has been experimentally shown that the Au solid-solubility [6] and diffusivity [7] in crystalline silicon ($c$-Si) is *lower* than that in amorphous silicon ($a$-Si). In the study of precipitation of implanted atoms, both the segregation and diffusion of gold atoms has been found to play a role: the segregation into a densely defected region where it exceeds the local solubility resulting in precipitation and the diffusion along the dislocations network until a node where again it exceeds the local threshold for precipitation [8].

Interesting results have been found as a result of the ion-nanostructure interactions [9-11]. Our recent experiments with high beam flux ($\approx 6.3 \times 10^{12}$ ions cm$^{-2}$ s$^{-1}$) showed a large mass transport from the nanostructured Au film into the Si which was absent under lower flux irradiation conditions. High resolution lattice image from the mixed regions showed the presence of gold silicide [12]. In the present work, we will report the effects of ion fluence under high flux conditions on the mass transport from nanostructured Au film into silicon substrate. Details about the phase separation of gold silicide and out diffusion of gold atoms to the surface regime are reported in this paper.

For the present work, 2.0 nm Au films were deposited using thermal evaporation by resistive heating method in high vacuum condition ($4 \times 10^{-6}$ mbar) on ≈2.0 nm thin native oxide covered Si(111) substrate. All irradiation experiment has been carried out at room temperature with 1.5 MeV $Au^{2+}$ ions at an impact angle of $5^0$ with a flux $6.3 \times 10^{12}$ ions $cm^{-2}$ $s^{-1}$ (corresponding to a current density of 2.0 µA $cm^{-2}$) and fluence values of $6 \times 10^{13}$ (F1), $1 \times 10^{14}$, (F2), $5 \times 10^{14}$ (F3) and $1 \times 10^{15}$ ions $cm^{-2}$ (F4). Transmission Electron Microscopy (TEM) measurements using the JEOL JEM-2010 were performed. High angle annular dark field (HAADF) imaging using the scanning TEM (STEM) was carried out using FEI Tecnai G2-F20 facility. Cross-sectional TEM (XTEM) samples were prepared by standard mechanical thinning followed by 3.5 keV Ar ion milling.

Figures 1(a) and (b) depicts representative of planar and bright field (BF) XTEM micrographs. Isolated nano Au islands on the Si surface were clearly evident from these figures. The average size and height of these isolated nanoislands were found to be 7.6±1.5 and 6.9±0.8 nm with 40 % surface coverage (we call this as a nanostructured Au film). Figures in 2(a) and (b) depicts the BF XTEM images of films irradiated at fluence of F1 and F2, respectively. The micrograph shown in Fig. 2(a) depicts a large mass transport extending up to a distance of about 60 nm from Au nanostructures into the substrate. Using the high resolution lattice images, it was found that the transported region had the composition of Au-Si Alloy ($Au_5Si_2$) [12]. The inset of Fgure 2(a) show a selected area electron diffraction (SAED) taken from the implanted region of the substrate (where no mixed layer was present). The SAED pattern clearly shows the amorphous nature of irradiated region. The enhanced diffusion (or transport) of gold atoms has been attributed to amorphous nature of the substrate silicon [12]. At a higher fluence value of F2 ($1 \times 10^{14}$), interestingly, a reduction in the maximum transported depth has been observed (maximum depth of ≈ 45 nm). This clearly shows the out diffusion of gold atoms from the mixed region towards the defect – rich surface region. The high resolution TEM image (taken from the rectangular region of Figure 2(b)) shown in Figure 2(c) confirms the formation of gold – silicide alloy composition (i.e., $Au_5Si_2$) from the experimentally determined lattice spacing of 0.305 ± 0.005 nm. This value matches with the (120) plane of hexagonal $Au_5Si_2$ phase [11, 12]. It should be noted that the surrounding silicon region was in amorphous in nature.

In the following, we discuss the mass transport phenomena at higher fluences, i.e. at $5\times10^{14}$ and $1\times10^{15}$ ions $cm^{-2}$. Figure 2(d) shows a BF – XTEM micrograph taken from the sample irradiated at a fluence of F3. At this fluence, crater formation on the surface has been observed. The average crater diameter found to be ≈ 25.5 nm with maximum and minimum diameters as 42.2 nm and 15.2 nm, respectively. The average depth of craters is ≈ 4.2 nm with maximum and minimum depth as 6.3 nm and 2.8 nm, respectively. Below the crater, the Au atoms appeared to be segregated to the surface, and the depth has reduced to an average value of ≈15.9 nm with maximum and minimum depths being 23.0 nm and 11.4 nm. At the fluences of F1 and F2, the maximum depths were found to be ≈ 60 and 45 nm, respectively and this has further reduced to a value of 23.0 nm at the fluence F3. Reduction of mixed region can be explained by considering the out diffusion of gold atoms to surface. The SAED pattern from the craters and absence of high resolution lattice image from the crater region reveal the amorphous nature of the crater material. It is highly possible to have an amorphous gold – silicide in the crater area. Figure 2(e) depicts a scanning TEM – high angle annular dark field (STEM – HAADF) image taken from the sample irradiated under similar conditions as shown in Figure 2(d). In this mode, the contribution from diffraction and phase contrast is significantly suppressed and the scattering cross-section of the electrons distributed in the annular area is roughly proportional to $Z^2$ (where Z is the atomic number). Due to this, the bright part in the image indicates the presence of heavy elements directly (Z-contrast imaging) [12]. STEM data confirms the higher Z presence, in our case gold atoms. When the fluence was increased to F4, we did not observe any noticeable morphological change from the sample that corresponding to F3 (comparable to Figure 2(d)). The craters were present on the surface with reduced depth (average depth is 12.1 nm with a maximum value of 18.6 nm). Au Sputtering from the surface at higher fluence mainly leads to this reduction in crater depth.

The contrast shown in Figures 2(a), 2(b) and 2(c), clearly indicates the out-diffusion of Au upon ion irradiation at higher fluences. It was already reported that Au (implanted species) does exhibit out-diffusion during thermal annealing processes [13 – 15]. In these reports, Au out-diffusion was explained as a result of recrystallization in Si upon annealing. It was observed that incorporated Au atoms (by means of implantation or by other means) use to find densely defect region at or near end of range which is supposed to be most densely defect region even after annealing. As a result, the implanted region recrystallizes and leaves out lot of defects at the end of range where Au atoms will be trapped and segregated. But in the present experimental observation, we did not observe recrystallization of the Si substrate even at fluence F4.

In many experimental observations, it was found that the ion beam mixing (IBM) is influenced by thermal diffusion of one species of atom to another. The magnitude and the sign of diffusion coefficient are important aspects for thermal diffusion assisted IBM. The sign of the effective diffusion coefficient can be either positive or negative. Positive coefficient leads to mixing between the components and negative coefficient leads to phase separation even though they formed alloy at low temperature. The sign of diffusion coefficient can be determined by the sign of $F$ factor [16] with the help of average cohesive energy ($|\Delta H_{coh}|$) and heat of mixing ($\Delta H_m$) [17]. For all the Au-Si phases, we got positive values of $\Delta H_m$ and negative values of $F$. For $Au_5Si_2$ (observed phase in the present experiment), $F$ factor is –0.38 (negative and non-zero magnitude which can not be neglected) indicating that the driving force for separation of the components in the ion-bombarded sample is pronounced. As a result, at higher fluence Au diffuse towards the defected rich surface region. The surface and the end of range ($a$-Si/$c$-Si interface) are known to be the more densely defect regions compared to the rest of the implanted regions. From our XTEM observation, the end of range has been found to be ≈ 520 nm, which is large compared to the maximum transported depth of about 60 nm. Hence, at higher flux, as the surface is nearer preferable destination for the in-diffused Au atoms, segregation as a result of out-diffusion has been observed at surface. The transient wafer temperature would also play a role in the segregation. The wafer temperature effects were discussed in ref [12].

In conclusion, the gold segregation to the defect rich surface regions has been observed at higher fluence without any additional thermal treatments.

**Acknowledgements:**

We thank B. Sundaravel and K. G. M. Nair (of IGCAR, Kalpakkam) for high flux irradiations and Chen and Liou for STEM measurements (of CCCM, Taipei) and Y. L. Wang for financial support of visit of P. V. Satyam to IAMS, Taipei.

**Figure captions:**

**Figure 1:** (a) and (b) correspond to the planar and cross-sectional micrographs for pristine 2.0 nm Au/Si system [12].

**Figure 2:** (a), (b), (d) and (f) are the bright field (BF) XTEM micrographs of the irradiated systems at fluence of $6\times10^{13}$, $1\times10^{14}$, $5\times10^{14}$ and $1\times10^{15}$ ions cm$^{-2}$, respectively. Inset of Figure 2(a) corresponds to the SAED pattern showing the amorphous nature of the implanted region of the substrate. Figure (c) corresponds to the high resolution TEM image of rectangular region of Figure 2(b). Figure (e) shows the STEM – HAADF image of the samples irradiated at a fluence of $5\times10^{14}$ ions cm$^{-2}$.

**Figure 1:**

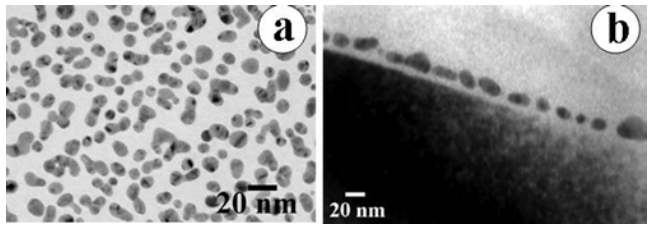

**Figure 2:**

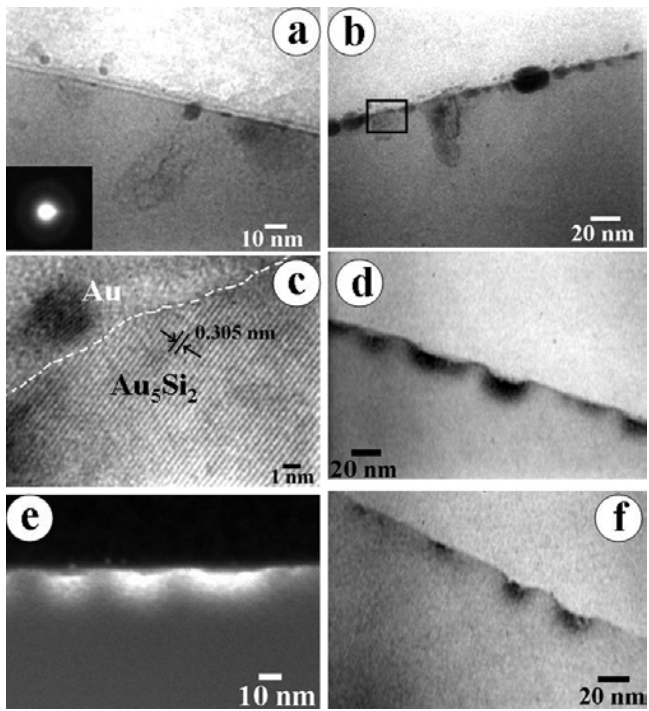